\documentstyle[12pt,aasms4]{article}

\lefthead{Wu et al.}
\righthead{SN 1006 Absorption Lines}

\begin{document}

\title{Far-UV Absorption Lines in the Remnant of SN 1006\altaffilmark{1}} 
\author{Chi-Chao Wu\altaffilmark{2}, D. Michael Crenshaw\altaffilmark{3},
Andrew J.S. Hamilton\altaffilmark{4}, Robert A. Fesen\altaffilmark{5}, 
Marvin Leventhal\altaffilmark{6}, and Craig L. Sarazin\altaffilmark{7}}

\altaffiltext{1}{Based on observations with the NASA/ESA Hubble Space
Telescope, obtained at the Space Telescope Science Institute, which is
operated by the Association of Universities for Research in Astronomy,
Inc., under NASA contract NAS5-26555.}
\altaffiltext{2}{Computer Sciences Corporation, Space Telescope Science 
Institute, 3700 San Martin Drive, Baltimore, MD 21218, wu@stsci.edu}
\altaffiltext{3}{Computer Sciences Corporation, Laboratory for Astronomy 
and Solar Physics, Code 681, NASA/Goddard Space Flight Center, Greenbelt, 
MD 20771}
\altaffiltext{4}{University of Colorado, JILA, Box 440, Boulder, CO 80309}
\altaffiltext{5}{Dartmouth College, Department of Physics and Astronomy, 
6127 Wilder Hall, Hanover, NH 03755}
\altaffiltext{6}{University of Maryland, Department of Astronomy, 
College Park, MD 20742-2421}
\altaffiltext{7}{University of Virginia, Department of Astronomy, P.O. Box
3818, Charlottesville, VA 22903}

\begin{abstract}
We have obtained a far-ultraviolet spectrum (1150 -- 1600 \AA) of a
hot subdwarf star behind the remnant of SN~1006 with the Faint Object
Spectrograph (FOS) on the {\it Hubble Space Telescope}. The
high-quality spectrum is used to test previous identifications of the
strong absorption features discovered with the {\it International
Ultraviolet Explorer}. These features have FWHM $=$ 4000 ($\pm$300) km
s$^{-1}$ and are {\it not} at the rest wavelengths of known
interstellar lines, as opposed to the broader ($\sim$8000 km s$^{-1}$
FWHM) \ion{Fe}{2} lines from the remnant centered at zero km s$^{-1}$
in near-UV FOS spectra. We confirm that the broad absorption features
are principally due to redshifted \ion{Si}{2}, \ion{Si}{3}, and
\ion{Si}{4} lines, which are centered at a radial velocity of 5100
($\pm$ 200) km s$^{-1}$. 

The \ion{Si}{2} $\lambda$1260.4 profile is asymmetric, with a nearly
flat core and sharp red wing, unlike the \ion{Si}{2} $\lambda$1526.7
and \ion{Si}{4} $\lambda\lambda$1393.8, 1402.8 profiles. One possible
explanation is additional absorption from another species. Previous
work has suggested that \ion{S}{2} $\lambda\lambda\lambda$ 1250.6,
1253.8, 1259.5 at a radial velocity of $\sim$6000 km~s$^{-1}$ is
responsible, but this would require a sulfur to silicon abundance
ratio that is at least a factor of ten higher than expected. Another
possible explanation is that the \ion{Si}{2} and \ion{Si}{4} profiles
are intrinsically different, but this does not explain the symmetric
(albeit weaker) \ion{Si}{2} $\lambda$1526.7 profile. 

\end{abstract}


\section{Introduction}

A hot subdwarf star (the ``SM star'') discovered by Schweizer \&
Middleditch (1980) lies on the far side of the remnant of SN~1006, and
provides an unusual opportunity to study its ejecta in absorption. IUE
observations (Wu et al. 1983; Fesen et al. 1988) revealed the presence
of strong, broad \ion{Fe}{2} absorption lines centered at zero
velocity. Subsequent FOS spectra (Wu et al. 1993)
were obtained at sufficient signal-to-noise and resolution to remove
the narrow interstellar and stellar lines, deblend the broad
\ion{Fe}{2} lines, and determine an intrinsic \ion{Fe}{2} profile that
is suitable for comparison with model predictions for Type Ia
supernovae. The \ion{Fe}{2} velocity profile is roughly symmetric
around zero km s$^{-1}$ and extends up to about $\pm$8000 km s$^{-1}$
at the continuum. The absorption-line width and angular size of the
remnant yield a lower limit to the distance of 1.9 kpc (more recent
estimates based on the NW optical filament's proper motion and shock
velocity yield 1.8 $\pm$ 0.3 kpc, see Laming et al. 1996). Whereas the
remnant contains approximately 0.014 $M_\odot$ of Fe$^{+}$, the
predicted mass of Fe from Type Ia supernova models (Nomoto,
Thielemann, \&\ Yokoi 1984; H\"{o}flich \& Khokhlov 1996) is about 25
times this value. However, most of the Fe in SNR 1006 could be in
higher ionization states than \ion{Fe}{2} (Hamilton \&\ Fesen 1988;
Blair, Long, \& Raymond 1996). 

The IUE observations also showed strong, broad absorption lines in the
far-UV, but these features were not at the rest wavelengths of known
interstellar lines. Wu et al. (1983) proposed that the outer portions
of the ejecta are less chemically processed. They identified the
features as \ion{Si}{2}, \ion{Si}{3}, and \ion{Si}{4} lines
originating from a clump in the far side of the ejecta, which is
moving at a radial velocity of $\sim$5000 km s$^{-1}$. Based on a
spectrum of SN 1006 with the {\it Hopkins Ultraviolet Telescope}
(HUT), Blair et al. (1996) attribute these lines entirely to Si.
However, Fesen et al. (1988) showed that the wavelength
correspondences are not exact, the line strengths are not quite in the
expected ratios, and some of the lines are asymmetric, and they
suggested that \ion{S}{2} at $\sim$6000 km s$^{-1}$ and \ion{O}{1} at
$\sim$6500 km s$^{-1}$ may also contribute to these features. Fesen et
al. (1988) suggested that these features may arise from clumps of
shocked material composed of intermediate mass elements in the far
side of the ejecta. We have obtained high signal-to-noise spectra of
the far UV lines to test these identifications, and we present the
most likely interpretations of these features. A companion paper
(Hamilton et al. 1996) presents a detailed physical picture of the
ejecta based on one of these interpretations. 

\section{Observations and Measurements}

FOS observations of the SM star behind SNR 1006 were obtained on 1993
September 17 -- 20, prior to the installation of COSTAR on HST,
through a circular aperture with a projected diameter of 1\arcsec
~(which gives instrumental profiles that are nearly Gaussian, see
Kinney 1992). The FOS/BLUE detector and G130H grating were used to
obtain a total integration time of 468 min over the range 1150 -
1600~\AA\ at a resolution of $\lambda$/$\Delta\lambda$ $\approx$ 1000.
The positions of the strong interstellar lines indicate that no
correction to the wavelength scale was needed. The absolute flux scale
of the G130H spectrum was adjusted by multiplying the fluxes by a
constant factor of 1.08, in order to match the FOS G190H fluxes in the
30 \AA\ region of overlap. The spectrum of the SM star was binned to
an interval of one-half resolution element ($\sim$0.7 \AA), for
comparison with standard star spectra (generously provided by Ralph
Bohlin). Based on photon counting statistics, the signal-to-noise
ratios per half-resolution element in the continuum of the SM star
spectrum are $\sim$30 at $\lambda$~$>$~1250~\AA\ and drop steadily
below 1250~\AA\ to $\sim$10 at 1150 \AA. 

Figure 1 shows the observed spectrum of the SM star behind SNR 1006,
and scaled spectra of hot standard stars. The spectra of HZ 44
(sdO) and BD+75\arcdeg 325 (sdO) are useful for identifying stellar
features, and the spectrum of BD+33\arcdeg 2642 (B2 IV) is
helpful for identifying interstellar features. Comparison of our FOS
spectrum with the HUT spectrum of the SM star (Blair, Long, \& Raymond
1996), obtained over the range 912 -- 1840~\AA, shows good agreement
at $\lambda$ $>$ 1300 \AA. At $\lambda$ $<$ 1300 \AA\, the spectra
begin to diverge, with the FOS continuum showing an excess of
$\sim$20\% over the HUT continuum at $\sim$1200 \AA. This may indicate
a problem with the FOS calibration at the shortest wavelengths;
further comparison of standard stars observed with HUT and FOS is
needed to investigate this problem. 

The observed FOS spectrum was dereddened using a value of $E_{B-V}$ =
0.1, which was determined from the 2200 \AA\ depression in the
continuum (Wu et al. 1993) and the interstellar extinction curve of
Savage \&\ Mathis (1979). The narrow interstellar and stellar
absorption lines were measured and removed as described in Wu et al.
(1993). All narrow lines were measured directly using a local
continuum; each line with an equivalent width greater than 3 times the
expected error was retained for further analysis. The narrow lines
were remeasured with a Gaussian fitting routine to determine better
positions (accurate to about $\pm$0.5 \AA), equivalent widths, and
errors in the equivalent widths, and were then removed from the FOS
spectrum by subtracting the appropriate Gaussian fits. Table 1 lists
the narrow lines that were measured. All of the strong interstellar
lines that are expected (Morton 1991) are seen; most of the remaining
narrow absorption features are difficult to identify unambiguously at
this resolution, but are very likely blends of lines that are common
in hot subdwarf stars, such as \ion{C}{3}, \ion{C}{4}, \ion{N}{3},
\ion{Fe}{4}, and \ion{Fe}{5} (Bruhweiler, Kondo, \&\ McCluskey 1981;
Dean \&\ Bruhweiler 1985). Each feature that has been measured is also
seen in at least one of the hot subdwarf spectra. 

Figure 1 shows that in addition to the narrow stellar and interstellar
lines in the spectrum of the SM star, the broad absorption features
from the SNR seen by Wu et al. (1983) and Fesen et al. (1988) are
present at the following approximate positions: 1217 \AA, 1282 \AA,
1331 \AA, 1421 \AA, and 1553 \AA. The position (i.e., centroid),
equivalent width, and velocity width (FWHM) of each of these features
were determined using a linear fit to local continuum points.
One-sigma errors in the measurements were determined from the photon
noise and different reasonable placements of the continuum and the
flux and wavelength extremes.  Table 2 gives these measurements and
errors. 

The positions and equivalent widths of the broad lines in Table 2 are
essentially the same as those in Wu et al. (1983) and Fesen et al.
(1988), who identify these features as principally redshifted Si
lines. The central radial velocities in Table 2 are calculated from
these identifications and the measured positions. The absorption
feature at 1217~\AA\ (see Figure 1) has contributions from geocoronal
Ly$\alpha$ emission, stellar and interstellar Ly$\alpha$ absorption,
redshifted \ion{Si}{2} $\lambda\lambda$ 1190.4, 1193.3 absorption, and
redshifted \ion{Si}{3} $\lambda$1206.5 absorption. Due to the blending
of these lines, no values are given for their widths and radial
velocities in Table 2. The 1331 \AA\ feature is weak and contaminated
by strong narrow interstellar lines, so the corresponding errors are
large. The width and radial velocity of each component of the
\ion{Si}{4} $\lambda\lambda$1393.8, 1402.8 doublet were derived using
the template profile fit described below. 

\section{Interpretation of the Broad Lines}

The broad absorption features in Table 2 are at the same radial
velocity (5100 $\pm$200 km s$^{-1}$) and have the same width (4000
$\pm$300 km~s$^{-1}$ FWHM, errors determined from the dispersion in
values and the individual measurement errors). Thus we confirm that
these features are primarily due to broad redshifted \ion{Si}{2},
\ion{Si}{3}, and \ion{Si}{4} lines. Further evidence based on the
strengths and profiles of the lines is presented below, along with
possible evidence of additional absorption from other species. 

Figure 2 shows expanded plots of the broad absorption features from
the dereddened spectrum, including the profiles prior to and after
removal of the narrow lines. The basic purpose of this plot is to
compare the strengths and profiles of the absorption features, to
determine if they can be interpreted in a consistent manner. The 1282
\AA\ feature represents the strongest \ion{Si}{2} line according to
Table 2, so its profile is adopted for comparison with the other
lines. This feature was normalized to the local continuum to create a
template profile, {\it assuming} that it is entirely due to redshifted
\ion{Si}{2} $\lambda$1260.4. The template was then reproduced at the
expected positions of the \ion{Si}{2} $\lambda$1304.4 and
$\lambda$1526.7 lines (at a radial velocity of v$_r$ $=$ 5100 km
s$^{-1}$) using the oscillator strengths for these lines (Morton
1991). The template was also reproduced at the expected positions of
the \ion{Si}{4} $\lambda\lambda$1393.8, 1402.8 and \ion{Si}{3}
$\lambda$1206.5 lines, but the scale factors were varied to provide
the best fit to the observed profiles. The scaled templates at the
local continuum levels are shown in Figure 2 for comparison with the
actual absorption features. 

First we consider the \ion{Si}{2} absorption lines, which should have
the same profiles. The close correspondence between the 1282 \AA\
template and the 1553 \AA\ feature, without the need for further
scaling, indicates that these features are indeed primarily due to
\ion{Si}{2} $\lambda$1260.4 and \ion{Si}{2} $\lambda$1526.7. There is
some evidence for additional broad absorption in the 1282 \AA\
feature, in the core and just redward of the core. Comparison of the
SM star spectrum with the hot subdwarf spectra in Figure 1 indicates
that there are no strong stellar or interstellar lines that could be
responsible for the excess broad absorption in the 1282 \AA\ feature.
The 1553 \AA\ feature has substantial contamination by narrow
lines, but tests using different scaling factors for their removal do
not result in a better agreement of the profiles. A straightforward 
interpretation is that the difference in profiles is due to additional
absorption in the 1282 \AA\ feature.

For the 1331 \AA\ feature, the scaled template is at least consistent
with the presence of \ion{Si}{2} $\lambda$1304.4 absorption, given the
noisiness of this feature. There may be additional absorption from
another species as well, even in excess of that present in the 1282
\AA\ template. However, the presence of strong narrow interstellar
lines in this region complicate the interpretation of this feature,
and it is possible that a larger contribution from the C I and C II
lines could explain more of the excess absorption. 

For the 1421 \AA\ feature, the template was reproduced at the
positions of the individual \ion{Si}{4} components and scaled to the
blue wing. There is also evidence for additional absorption in the
1282 \AA\ feature, in the same location as noted before. However,
another possibility is that the \ion{Si}{2} and \ion{Si}{4} profiles
are intrinsically different (Hamilton et al. 1996).

The complicated nature of the 1217 \AA\ feature makes it difficult to
obtain an accurate deconvolution. However, the \ion{Si}{3} line is
sufficiently distinct to provide an estimate of its strength by
fitting the red wing of the feature. The poor match of the template to
the blue wing of this feature is a concern; it may be due to an
error in the flux calibration in this region, as discussed earlier. 

We conclude that previous work is correct in the identification of
\ion{Si}{2}, \ion{Si}{3}, and \ion{Si}{4} as the main contributors to
the broad absorption lines in the far-UV. In addition, these lines
have the same approximate radial velocities and widths, despite the
range in ionization level. However, the profiles of the features
attributed to \ion{Si}{2} are dissimilar.

As noted by Fesen et al. (1988), a possible explanation for the
different Si II profiles is the presence of additional absorption in
the 1282 \AA\ and 1331 \AA\ features, and they suggest that \ion{S}{2}
$\lambda\lambda\lambda$ 1250.6, 1253.8, 1259.5 and \ion{O}{1}
$\lambda$ 1303.2 lines may be responsible. The positions and widths
needed for these lines to reproduce the observed features was
determined from an estimate of the excess absorption. The
corresponding radial velocities are noted in Figure 2. If the residual
absorptions are interpreted as \ion{S}{2} and \ion{O}{1} lines, they
are at the same approximate radial velocity ($\sim$6800 km s$^{-1}$),
which is substantially higher than that of the Si lines. Although the
residuals are noisy, they are clearly narrower than the Si lines
($\sim$1800 km s$^{-1}$). The next section points out the difficulties
in this interpretation. 

\section{Discussion}

An interesting result from the FOS observations is the large intrinsic
width of the far-UV absorption lines from SNR 1006, which might be
expected to resolve into individual components if the absorption
arises from small knots of gas. In fact, the profiles of the deblended
Si lines start at v$_{r}$ $=$ 1500 ($\pm$500) km s$^{-1}$ and extend
out to v$_{r}$ $=$ 8500 ($\pm$500) km s$^{-1}$, which is close to the
extent of the red wing of the \ion{Fe}{2} lines (at $\sim$8000 km
s$^{-1}$). This suggests that the Si lines do not arise from a small
clump or a few clumps of material, but rather in a layer of gas in
bulk motion in the far side of the ejecta, and that Si and Fe may be
intermixed on the far side. We note that there is substantial
observational and theoretical evidence for mixing of the interior and
exterior layers in a Type II supernova (SN1987A, see Bussard, Burrows,
\&\ The 1989; Yamada \&\ Sato 1991). 

The equivalent widths of the broad absorption lines have not varied at
the $\geq$20\% level over $\sim$12 years of UV observation, since the
first IUE observations by Wu et al. (1983). This lack of variability
is also difficult to reconcile with small clumps, which have
significant transverse velocities because the hot subdwarf is
$\sim$2.5$'$ from the center of the SNR. From the 12-year time
interval, projected offset of the subdwarf, and observed velocities, a
lower limit of 0.02 pc is obtained for the size of the absorbing
region perpendicular to the line of sight. 

Another interesting result from the FOS spectra is that there is no
evidence for strong blue-shifted counterparts to the redshifted Si
lines (although there appears to be a weak depression in the $\sim$
1375 \AA\ region, just blueward of the \ion{Si}{4}
$\lambda\lambda$1393.8 line). The lack of blue-shifted Si lines can be
explained by asymmetry in the original supernova explosion and/or
asymmetry in the distribution of the interstellar medium surrounding
the supernova (along with a significant interaction of the ejecta with
the interstellar medium). Hamilton et al. (1996) discuss these
possibilities in detail, and argue for a much lower density of the
interstellar medium on the far side of the remnant compared to the
rest of the remnant. 

The apparent excess absorptions in the 1282 \AA\ and 1331 \AA\
features are difficult to understand. They may be interpreted as
\ion{S}{2} and \ion{O}{1} lines with centroids at higher redshifted
radial velocities, but then the relative abundance of sulfur would
have to be much higher than predicted from models of Type Ia
supernovae. The column densities of the Si ions, determined from
direct integration of the optical depths across the profiles, are
N(\ion{Si}{2}) $=$ 6.7 ($\pm$0.6) x 10$^{14}$ cm$^{-2}$,
N(\ion{Si}{3}) $=$ 3.1 ($\pm$0.6) x 10$^{14}$ cm$^{-2}$,
and N(\ion{Si}{4}) $=$ 3.6 ($\pm$0.5) x 10$^{14}$ cm$^{-2}$.
(The \ion{Si}{2} value comes from \ion{Si}{2} $\lambda$1526.7 at 1553
\AA, whereas Hamilton et al. [1996] obtain a higher value by assuming
the entire 1282 \AA\ feature is due to \ion{Si}{2} $\lambda$1260.4.)
The column densities of the \ion{S}{2} and \ion{O}{1} lines, estimated
from the excess absorption, can be used to determine a lower limit to
their abundance {\it relative} to the total Si abundance (assuming no
other ionization states for Si). The ratio of oxygen to silicon
abundance is $\geq$ 1, which is close to the expected ratio of 1 -- 2
(Nomoto et al. 1984). The ratio of sulfur to silicon abundance is
$\geq$5.5, which is a factor of ten higher than the expected ratio of
$\sim$0.5 (Nomoto et al. 1984). Thus, although the identification of
the silicon lines is secure, the interpretation of excess absorption
due to redshifted sulfur and oxygen lines is in doubt. 

Hamilton et al. (1996) discuss an alternate interpretation of the 1282
\AA\ feature, in which the asymmetric profile is due to shocked and
unshocked components of \ion{Si}{2} $\lambda$1260.4 absorption.
However, this does not explain the difference between the observed
\ion{Si}{2} $\lambda$1260.4 and \ion{Si}{2} $\lambda$1526.7 line
profiles. Since the present FOS data are insufficient to resolve this
apparent discrepancy, observations at higher resolution (to fit and
remove the narrow lines more accurately) and higher signal-to-noise
(to get better profiles of the weak broad lines) are desirable.
Improved data will also set stricter limits on blueshifted absorption
features from the remnant's approaching hemisphere. Similar
observations of a hot subdwarf star that is chosen to match the
effective temperature and surface gravity of the SM star would also be
helpful in determining an accurate continuum and removing stellar
absorption features. 

\acknowledgments
This work was supported by STScI grant GO-3621.01-91A to the Computer
Sciences Corporation. We thank the referee for helpful comments.

\clearpage

\begin{deluxetable}{ccclcccl}
\tablecolumns{8}
\footnotesize
\baselineskip 15pt
\tablecaption{Narrow absorption-line measurements. \label{tbl-1}}
\tablewidth{0pt}
\tablehead{
\colhead{$\lambda$} & \colhead{E.W.}   &                
\colhead{$\sigma$$_{E.W.}$} & \colhead{Identification} &
\colhead{$\lambda$} & \colhead{E.W.}   &                
\colhead{$\sigma$$_{E.W.}$} & \colhead{Identification} \\
\colhead{(\AA)} & \colhead{(\AA)}   & \colhead{(\AA)} & \colhead{} &
\colhead{(\AA)} & \colhead{(\AA)}   & \colhead{(\AA)} & \colhead{}  
}
\startdata
1175.7 &1.14 &0.19 &\ion{C}{3} $\lambda$1175.7 (stellar)  &1328.4 &0.23 &0.09 &\ion{C}{1} $\lambda$1328.8 \nl  
1183.8 &0.44 &0.12 &                                      &1334.6 &0.53 &0.13 &\ion{C}{2} $\lambda$1334.5 \nl  
1189.8 &0.78 &0.12 &\ion{Si}{2} $\lambda$1190.4           &1454.9 &0.21 &0.05 & \nl                            
1193.5 &0.66 &0.08 &\ion{Si}{2} $\lambda$1193.3           &1460.5 &0.16 &0.06 & \nl                            
1200.2 &1.22 &0.10 &\ion{N}{1} $\lambda$1200              &1466.7 &0.49 &0.13 & \nl                            
1206.7 &0.55 &0.11 &\ion{Si}{3} $\lambda$1206.5           &1499.1 &0.41 &0.10 & \nl                            
1229.5 &0.27 &0.11 &                                      &1505.3 &0.22 &0.06 & \nl                            
1239.4 &0.23 &0.07 &                                      &1526.6 &0.61 &0.08 &\ion{Si}{2} $\lambda$1526.7 \nl 
1247.4 &0.18 &0.05 &                                      &1534.1 &0.31 &0.07 & \nl                            
1250.6 &0.30 &0.10 &\ion{S}{2} $\lambda$1250.6            &1538.5 &0.16 &0.07 & \nl                            
1253.4 &0.52 &0.11 &\ion{S}{2} $\lambda$1253.8            &1548.4 &0.34 &0.07 &\ion{C}{4} $\lambda$1548.2 \nl  
1260.0 &1.06 &0.12 &\ion{Si}{2} $\lambda$1260.4           &1550.8 &0.29 &0.06 &\ion{C}{4} $\lambda$1550.8 \nl  
1296.4 &0.18 &0.06 &\ion{S}{1} $\lambda$1295.8            &1560.5 &0.25 &0.06 &\ion{C}{1} $\lambda$1560.3 \nl  
1301.8 &0.13 &0.05 &\ion{O}{1} $\lambda$1302.2            &1582.7 &0.31 &0.08 & \nl                            
1304.2 &0.13 &0.08 &\ion{Si}{2} $\lambda$1304.4           & \nl
\enddata
\end{deluxetable}

\clearpage
\begin{deluxetable}{crclc}
\tablecolumns{5}
\footnotesize
\tablecaption{Broad far-UV absorption lines in SNR~1006. \label{tbl-2}}
\tablewidth{0pt}
\tablehead{
\colhead{$\lambda$} & \colhead{E.W.} & \colhead{FWHM} 
& \colhead{Identification} & \colhead{v$_{r}$}\\
\colhead{(\AA)} & \colhead{(\AA)} & \colhead{(km s$^{-1}$)}
& \colhead{} &\colhead{(km s$^{-1}$)} 
}
\startdata
1217.0$\pm$1.0 & 20.9$\pm$1.7 & --------
& \ion{Si}{2} $\lambda\lambda$1190.4, 1193.3; \ion{Si}{3} $\lambda$1206.5;
stellar L$\alpha$ & -------- \nl

1282.3$\pm$0.4 & 10.5$\pm$0.6 & 3900$\pm$200
& \ion{Si}{2} $\lambda$1260.4; residual absorption? & 5200$\pm$100  \nl

1330.6$\pm$3.0 &  1.7$\pm$0.4 & 3400$\pm$900
& \ion{Si}{2} $\lambda$1304.4; residual absorption? & 6000$\pm$700 \nl

1421.2$\pm$0.4 &  4.9$\pm$0.7 & 4000$\pm$300
& \ion{Si}{4} $\lambda\lambda$1393.8, 1402.8 & 4900$\pm$100 \nl

1552.9$\pm$0.5 &  3.2$\pm$0.3 & 4200$\pm$200
& \ion{Si}{2} $\lambda$1526.7 & 5100$\pm$100 \nl

\enddata
\end{deluxetable}

\clearpage

\figcaption[fig1.ps]{Observed FOS G130H spectra of the SM star
behind SNR 1006 and standard stars HZ 44 (sdO), BD+75\arcdeg 325 (sdO),
and BD+33\arcdeg 2642 (B2 IV). Each standard star spectrum has been
scaled to the same average flux as the dereddened S-M spectrum, and
subsequently offset by adding a constant positive flux. \label{fig1}} 

\figcaption[fig2.ps]{Expanded plots of the dereddened FOS spectrum
of the SM star. The dotted and solid lines give the spectra prior to and
after the removal of the narrow absorption lines, respectively. The
smooth solid lines are the scaled 1282 \AA\ template at the
expected positions of the redshifted \ion{Si}{2}, \ion{Si}{3}, and
\ion{Si}{4} lines. Narrow interstellar lines and broad SNR lines are
labeled below and above the spectra, respectively (question marks 
indicate uncertain identifications). \label{fig2}} 

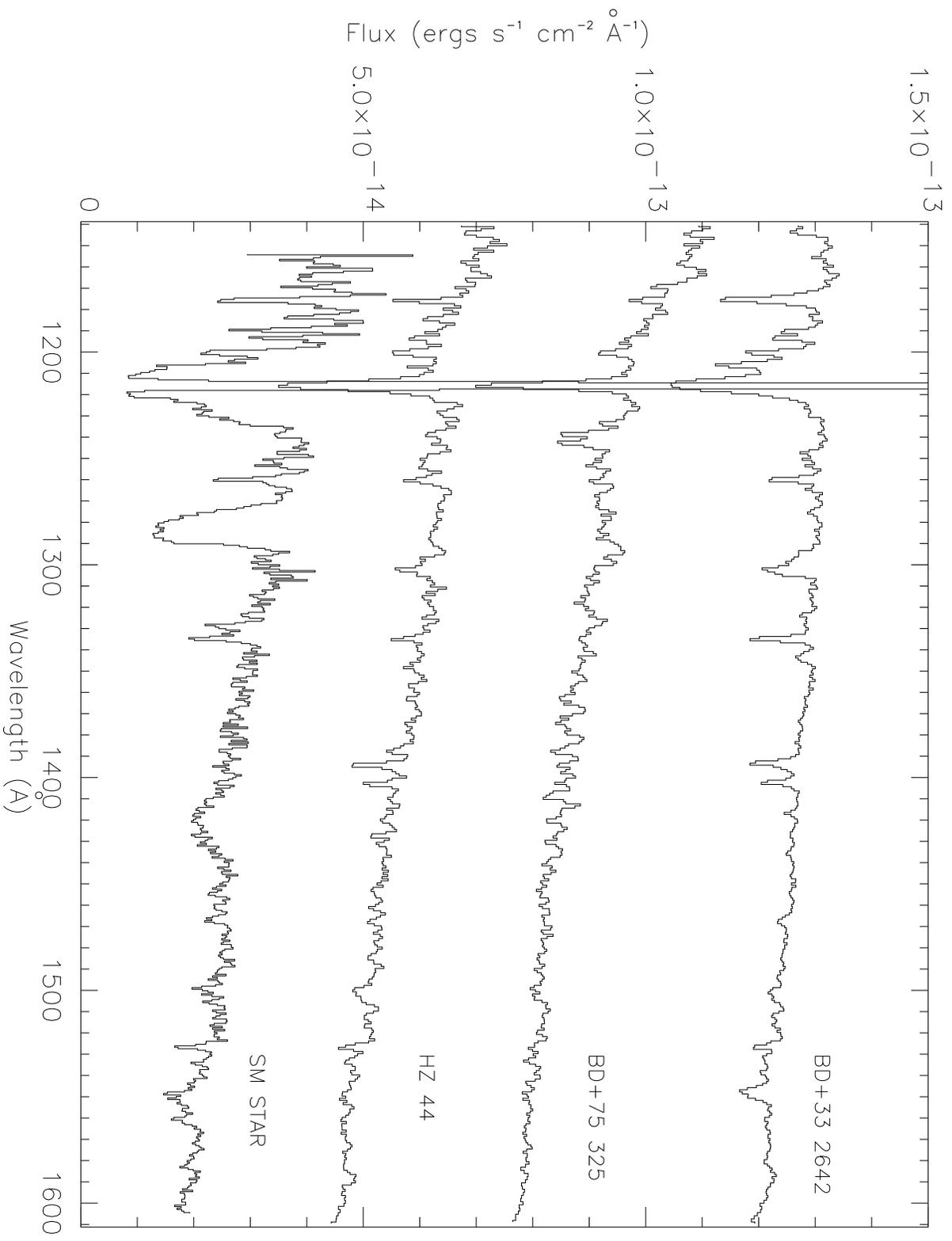
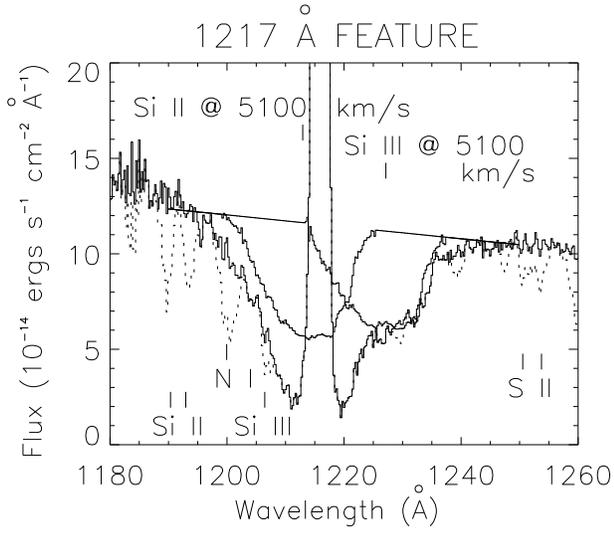
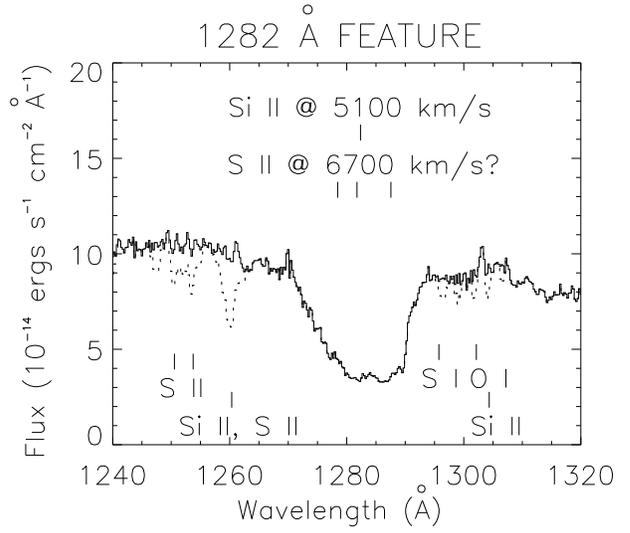
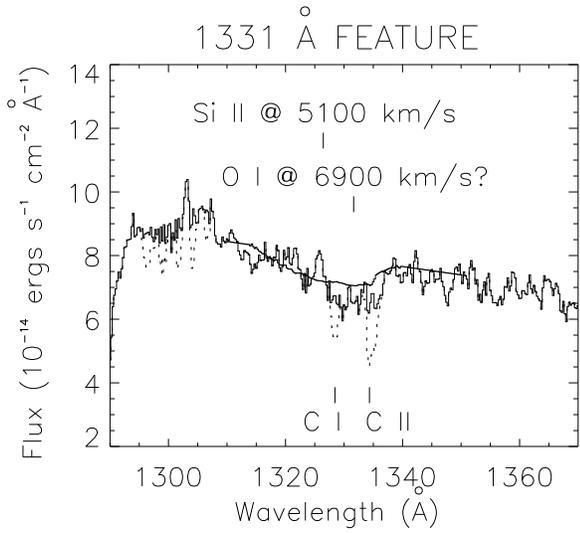
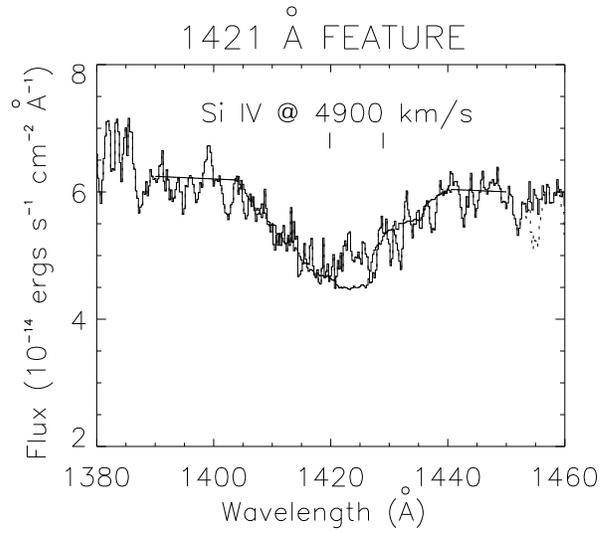
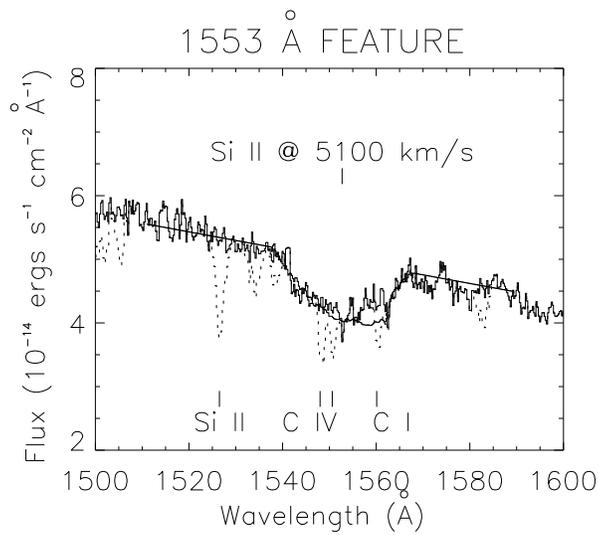
\end{document}